\documentclass[10pt,preprint2]{aastex}

\def \apj {ApJ}

\def \apjl {ApJ}
\def \solphys {Solar Phys.}
\def \pasj {Pub. Astron. Soc. Japan}
\def \aap {A\&A}

\newcommand{\citeN}[1]{\citeauthor{#1} (\citeyear{#1})}
\newcommand{\citeNP}[1]{\citeauthor{#1} \citeyear{#1}}

\newcommand{\CaII}{\ion{Ca}{2}}

\newcommand{\HeI}{\ion{He}{1}}

\shortauthors{Socas-Navarro et al}
\shorttitle{Fine Structure in the Umbral Chromosphere}

\begin{document}

\title{Direct Imaging of Fine Structure in the Chromosphere of a Sunspot
Umbra}

\author{H. Socas-Navarro}
   	\affil{Instituto de Astrof\'\i sica de Canarias, Avda V\'\i a
          L\' actea S/N, La Laguna 38200, Tenerife, SPAIN}
	\email{hsocas@iac.es}

\author{S. W. McIntosh, R. Centeno, A. G. de Wijn, B. W. Lites} 
   	\affil{High Altitude Observatory, NCAR\thanks{The National Center
	for Atmospheric Research (NCAR) is sponsored by the National Science
	Foundation.}, 3080 Center Green Dr, Boulder, CO 80301, USA}




\date{}%

   
               
\begin{abstract}
  High-resolution imaging observations from the Hinode spacecraft in
  the \CaII~H line are employed to study the dynamics of the
  chromosphere above a sunspot. We find that umbral flashes and other
  brightenings produced by the oscillation are extremely rich in fine
  structure, even beyond the resolving limit of our observations
  (0.22''). The umbra is tremendously dynamic, to the point that our
  time cadence of 20~s does not suffice to resolve the fast lateral
  (probably apparent) motion of the emission source. Some bright
  elements in our dataset move with horizontal propagation speeds of
  30~km~s$^{-1}$. We have detected filamentary structures inside the
  umbra (some of which have a horizontal extension of $\sim$1500~km)
  which, to our best knowledge, had not been reported before. The power
  spectra of the intensity fluctuations reveals a few distinct areas
  with different properties within the umbra that seem to correspond
  with the umbral cores that form it. Inside each one of these areas
  the dominant frequencies of the oscillation are coherent,
  but they vary considerably from one core to another. 
\end{abstract}


\section{Introduction}

The umbra of sunspots has been traditionally regarded as a relatively
calm and well understood place where nothing terribly exciting takes
place. Contributing to this idea is the contrast with the surrounding
penumbra, whose nature, dynamics and structure are still rather
mysterious. A first glance at some of the recent reviews on sunspots
shows that they are mostly focused on the problems posed by the
penumbra, sometimes with no mention at all of the umbra (e.g.,
\citeNP{scharmeretalreview}; \citeNP{bellotreview}). This is mainly
because the observations are in good agreement with the prevailing
paradigm of what a sunspot looks like. The idea is that a sunspot can
be viewed as the cross-section of a thick flux tube with magnetic
field lines that are vertical at the center and then fan out
radially. 
Of course this picture is too simplistic to
explain what is seen in the penumbra but it does a good job in terms
of the umbra. There is still the issue of umbral dots and light
bridges but those can be addressed by considering magneto-convection in
the picture (\citeNP{magnetoconvection}).







However, the umbra looks very different when we observe its
chromosphere. The photospheric oscillation becomes steeper as the wave
travels upward into a lower density regime before forming
a shock. Sometimes the hot shocked material produces clearly visible
emissions in the cores of chromospheric lines
known as umbral flashes. First discovered by 
\citeN{BT69} (see also \citeNP{W69}), their study is still a hot topic
of research (\citeNP{UF1}; \citeNP{UF2}; \citeNP{VdV03}; \citeNP{UF4};
\citeNP{UF5}; \citeNP{sci}; \citeNP{UF7}; \citeNP{UF8}). 
The flashes are seen as bright patches with typical
diameters of a few arc-seconds. The prevailing paradigm views the
umbral chromosphere as very homogeneous in the horizontal
direction. This is not only because the umbral flashes are relatively
large in extension but also because the strong magnetic field, which
dominates the dynamics in the low-$\beta$ regime, must relax itself to
an equilibrium configuration similar to the cartoon picture mentioned
above with field lines smoothly fanning out.


The first indication that something more complicated was going on came
with the analysis of polarization observations of 
\citeN{sci} (see also 2000a)\nocite{SNRCTB01}
. They detected the occurrence of ``anomalous''
Stokes spectra in coincidence with the peak of the oscillation. The
anomalous Stokes profiles occurred periodically and in coincidence
with the umbral flashes when they were visible\footnote{ Sometimes the
  emission was too weak to be recognized as a true ``flash'' but the
  formation of an anomalous polarization profile would still indicate
  that the wave front had reached the chromosphere.}.  After ruling
out other scenarios, the authors concluded that such unusual profiles
had to be produced by the coexistence of two different components
within the resolution element ($\sim$1''). One of the components was
active, harboring very hot up-flowing material and giving rise to the
observed emission, while the other component was cool and at rest. The
notion that the chromospheric oscillation has very fine structure
would introduce important complications in a problem that was thought
to be reasonably well understood and the whole idea was received
without much enthusiasm by the community.

In fact, the only 
other paper published thus far that deals with this issue is that of
\citeN{Cetal} in which a second independent observation points in the
direction of fine structure inside the emission patch. In this work
the authors had observed the infrared \HeI \ multiplet at 1080.3~nm
and their interpretation also leads to the coexistence of an active
and a quiet component in the resolution element ($\sim$1''). 





Imaging observations at La Palma with the SST and the DOT by
\citeN{VdV03} start to reveal some small bright elements in the umbra,
although with their spatial resolution the umbral flashes still
appeared as a monolithic structure. In fact, the authors did not pay
attention to that issue in the paper. Here we use similar observations
from the Hinode (\citeNP{Hinode1}; \citeNP{Hinode4})
spacecraft (but with improved spatial resolution) that 
allow us to resolve for the first time an intermixed pattern of bright
and dark features in the oscillation elements, even inside the umbral
flashes themselves.



\begin{figure*}
\epsscale{2}
\plotone{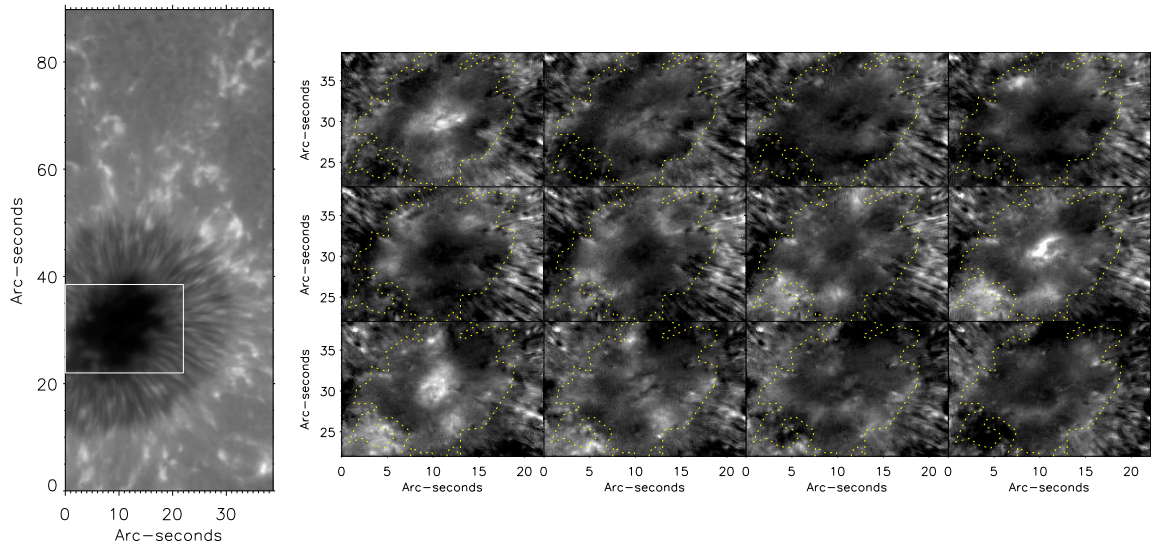}
\caption{Left: Time average of the intensity in the observed field of
  view. The area enclosed in the white rectangle is displayed in the
  right panel. Right: 12 successive snapshots (running from top left
  to bottom right) of the intensity after subtracting a 20-frame
  running mean and saturating the scale to show the umbral
  emission. The yellow dotted lines show the umbra-penumbra boundary
  as a reference.
\label{fig:sequence}
}
\end{figure*}

\section{Observations and data reduction}

The observations were taken on 2008 March 29$^{th}$ at 11:53 UT using
the broadband filter of the Solar Optical Telescope onboard the Hinode
spacecraft. For details on the instrumentation see \citeN{hinode}. We
observed time series of the sunspot in active region number
NOAA~10988, located at coordinates (-501'',-104'') from disk
center. The time cadence was 20~s. The SOT broadband filter for the
\CaII \ H line has a pass-band of 0.3~nm. Standard reduction
procedures, including flatfield and dark current correction, were
applied. The time span of our dataset is of approximately 1 hour and 4
minutes. The original series contains some bad frames (i.e., frames
with no data in all or part of the field of view) which we avoid by
using only a smaller subset of 40 minutes. In this manner we work with
a continuous good data set of homogeneous quality and regular
sampling. The diffraction limit of the telescope at this wavelength is
approximately 0.2'', which is only slightly under-sampled by the pixel
size (after binning) of 0.11''. Therefore, our spatial resolution is
0.22'' and this figure is of course constant through the entire
series.


\section{Results}

The field of view of our observation can be seen in
Figure~\ref{fig:sequence} (left panel). It encompasses most of the
sunspot and some of the surrounding moat. In this paper we shall focus
on the umbra, in particular the sub-field indicated by the white
rectangle in the figure. The right panel of the figure shows a
sequence of successive snapshots to give the reader an idea of the
complex and intricate patterns found, as well as their very vigorous
dynamics. In order to enhance the contrast and present the oscillatory
behavior clearly in this and the following figures, we have subtracted
a running mean from each snapshot. The mean that we subtract is
composed of 20 frames (spanning 400 seconds) around the snapshot under
consideration. This procedure is very similar to the one used by
\citeN{VdV03}. The umbra-penumbra boundary is overlaid in yellow
(dotted line) for reference. The series presented in the figure shows
12 successive snapshots, from top left to bottom right, covering 4
minutes of real time. The dominant oscillation period in the umbra is
approximately 3 minutes (see below).

The figure shows some hints of the complex behavior of the umbral
dynamics which are more clearly appreciated by viewing the data as a
movie. We have included it as an mpeg file with the electronic version
of this paper. Umbral flashes are visible occasionally at different
locations, some times near the center of the umbra and others near the
umbra-penumbra boundary. A more diffuse (but still structured) bright
component can be seen moving very rapidly. These motions are unlikely
to be actual plasma motions because they would be unrealistically fast
($\sim$100~km~s$^{-1}$) and the associated Doppler shifts would have
been observed already in observations away from the disk center. Most
likely the apparent motion is simply caused by the wavefront reaching
the line formation height at different times in different locations.

The changes in the diffuse bright component are so fast that they are
under-sampled by our 20~s cadence. It is then impossible to track a
feature as it moves around because the whole structure has changed
from one snapshot to the next. When one watches the movies, the eye
tends to interpret the motion as a wave moving around and bouncing off
of the umbra-penumbra boundary or the relatively bright structures
that protrude into the umbra. However, given that the presumed
propagation is under-sampled in time, this might just be an
illusion. This component is probably the same that \citeN{VdV03}
describe as arches emanating radially from the flashes. However, we do
not see that morphology or behavior in our data. Perhaps the
differences in spatial resolution (better in our case) or the time
sampling (better in their case) are responsible for the
discrepancy. In any case, one should be careful not to over-interpret
the sequence of images.


The visual appearance of the movie is very chaotic in the umbra
whereas the penumbra exhibits a much smoother behavior. Very fine
penumbral filaments are clearly visible in the data, swaying with a
slow and smooth oscillation pattern (probably running penumbral waves,
described by \citeNP{NT74}). When umbral flashes occur, they exhibit
fine structure. They are actually made of a mixture of bright and dark
dots. 

\begin{figure*}
\epsscale{2}
\plotone{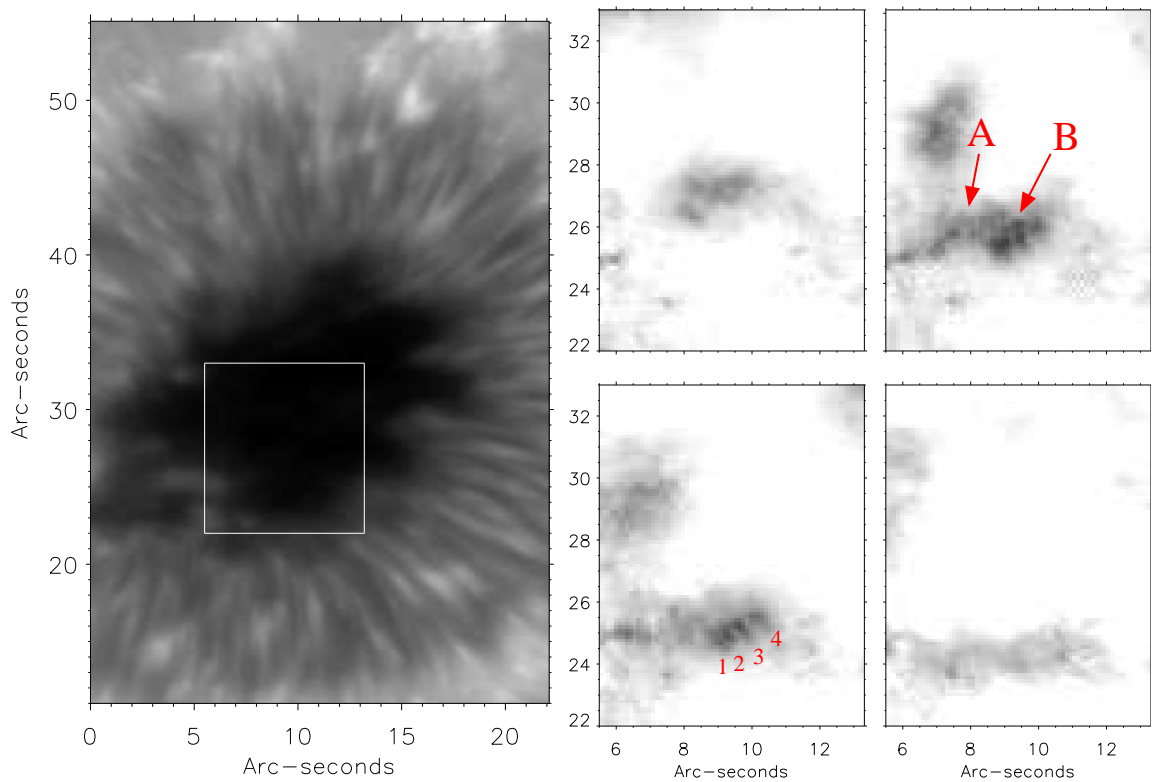}
\caption{Left: Field of view analyzed in the right panel. Right: 4 successive
  snapshots showing the propagation of a wavefront inside a
  filamentary structure. A 20-frame running median has been subtracted
  from the data. The grey scale has been inverted to show a
  negative for better viewing. Dark structures in the figure are
  actually bright and the white background is actually dark.
\label{fig:filaments}
}
\end{figure*}

\begin{figure*}
\epsscale{2}
\plotone{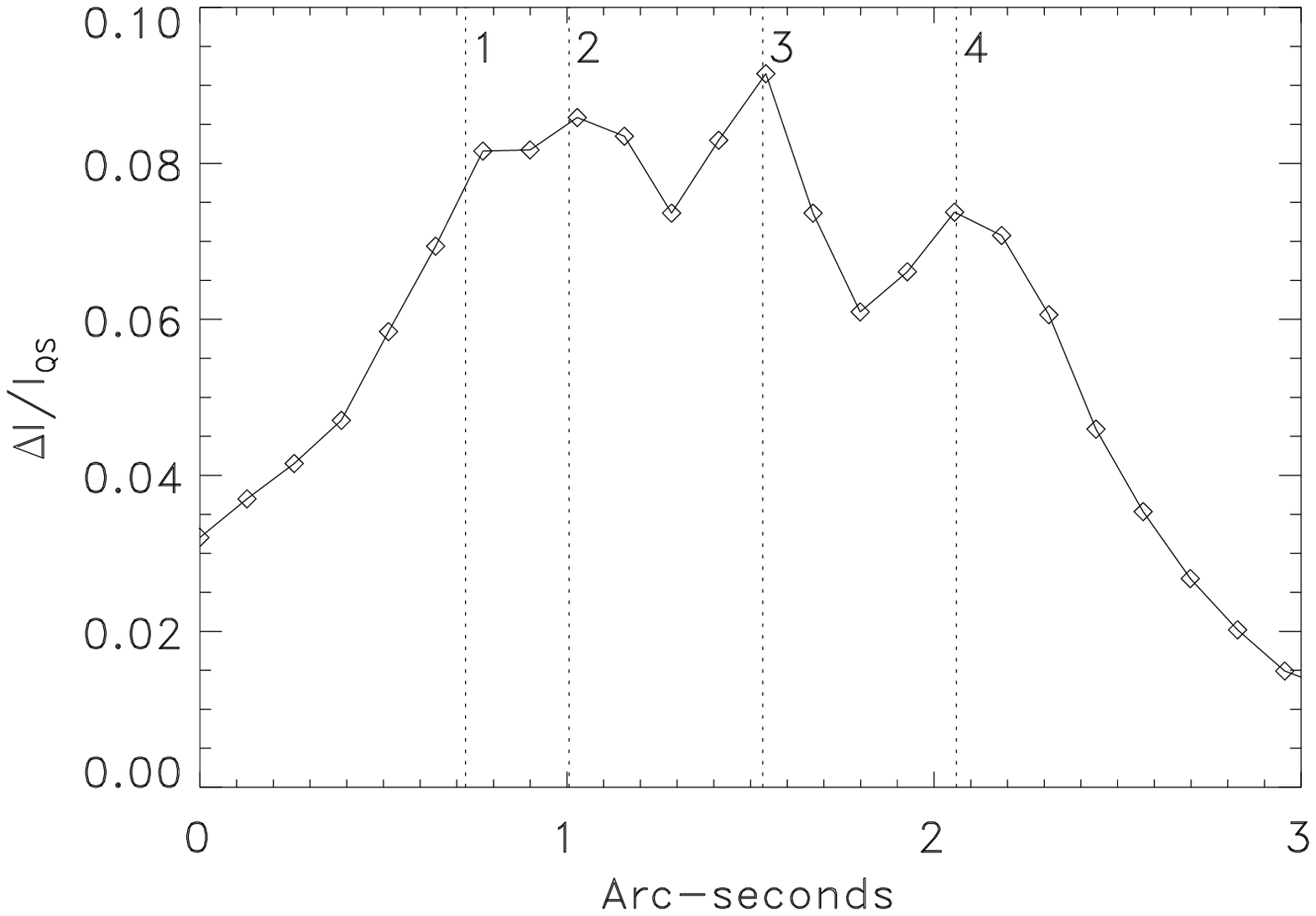}
\caption{Cut perpendicular to the direction defined by the filaments
  marked with arrow ``B'' in Figure~2. The curve shows the intensity
  after subracting a 20-frame median in units of the quiet Sun average
  continuum. The positions of the four filaments labeled with numbers
  in Figure 2 are shown with vertical dotted lines.
\label{fig:filamentcut}
}
\end{figure*}

Sometimes, however, the bright emission of a flash exhibits
filamentary structure. An example is given in Fig~\ref{fig:filaments}
(right), where one such structure (marked with arrow ``B'') appears in
the second frame (top right), remains in the third (bottom left)
although the brightening has moved towards the bottom of the figure,
and finally has already disappeared by the fourth frame (bottom
right). Actually, this figure shows two distinct filament
systems. One, around coordinates (8'',25''), marked with arrow ``A''
in the figure, is nearly oriented along the $y$-axis and is extremely
thin. In fact, it is barely resolved in our data and is likely to have
even finer structure. The other one, mentioned above, is around
coordinates (9'',26'') and marked with arrow ``B'' in the figure. This
second system is more prominent and shows four thick filaments that
are still visible in the third frame (identified by the numbers next
to them). Figure~\ref{fig:filamentcut} shows a cross plot with the
intensity fluctuation along a line perpendicular to the filaments. We
can see that the contrast is not very high, except for filament number
3, which makes it rather hard to discern the structure in the
grayscale panels.

We computed the horizontal speed of the brightenings in the four thick
filaments as they moved from the second to the third frame of
Figure~\ref{fig:filaments}, obtaining (from 1 to 4) the following
values in km~s$^{-1}$: 23$\pm$4, 20$\pm$2, 33$\pm$1 and 27$\pm$3. The
uncertainty is estimated as the difference in speeds obtained for
different points along the same filament. Some of the speeds obtained
differ in a statistically significant manner, which may be an
indication of differences in field geometry and, possibly, in the wave
propagation paths. The speeds are much larger than the vertical
Doppler velocities observed in umbral flashes (\citeNP{BT69};
\citeNP{SN01}) so it is unlikely that we are seeing an actual motion
of plasma along the field lines.
 
The presence of filaments whose geometry has a significant horizontal
component inside the umbra is certainly puzzling. The filament system
seen in Fig~\ref{fig:filaments} has a horizontal extension of about
2,000~km. Even taking a conservatively large range for the formation
height of the Ca~H line core of 1,000~km (the response functions of
\citeNP{Carlsson} in the \citeNP{FAL} C model would indicate more like
$\sim$500~km), the filament orientation must be nearly
horizontal. Spanning 2,000~km in the horizontal direction and less
than 1,000~km in the vertical, they would deviate less than 25$^{\rm
  o}$ from the horizontal. However, we do not make a strong claim in
this sense because one has to be careful with such estimates as the
formation height is only a meaningful concept in a plane-parallel 1D
atmospheric model.

\begin{figure*}
\epsscale{2}
\plotone{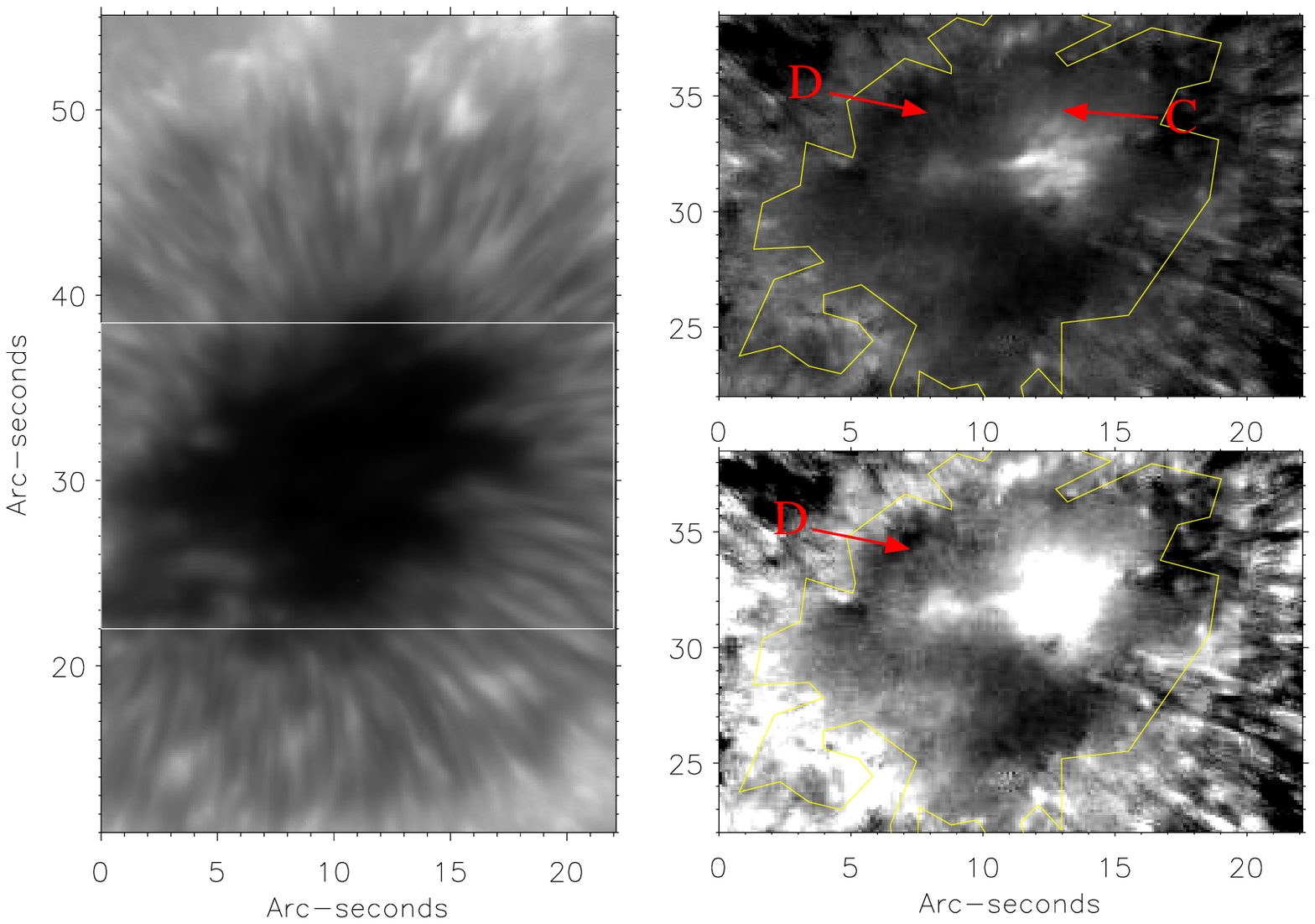}
\caption{Left: Field of view analyzed in the right panel. Right:
  Snapshot showing filamentary structure around an umbral flash. A
  20-frame running median has been subtracted from the data. The
  top and bottom images are the same but with different levels of
  saturation of the grey scale to show different details.
\label{fig:filaments2}
}
\end{figure*}

This filamentary system is seen persistently at other times, whenever
a strong brightening occurs at this location (for instance, see frames
number 32, 49, 56 or 156 in the movie). It is almost as if the
umbral flashes acted as a spotlight that moves around the dark umbra
and allows us to see things as it illuminates here and
there. Figure~\ref{fig:filaments2} shows a different filamentary
system. The top right figure has been saturated to show in detail the
structure of the umbral flash and the filaments that emanate almost
radially from the top (pointed by arrow ``C''). The bottom right
figure has been enhanced to show the filaments emanating from the
weaker brightening to the left of the main one (pointed by arrow
``D''). These weaker filaments appear to be oriented radially from a
common center at coordinates (8'', 32'') approximately.

\begin{figure*}
\epsscale{2}
\plotone{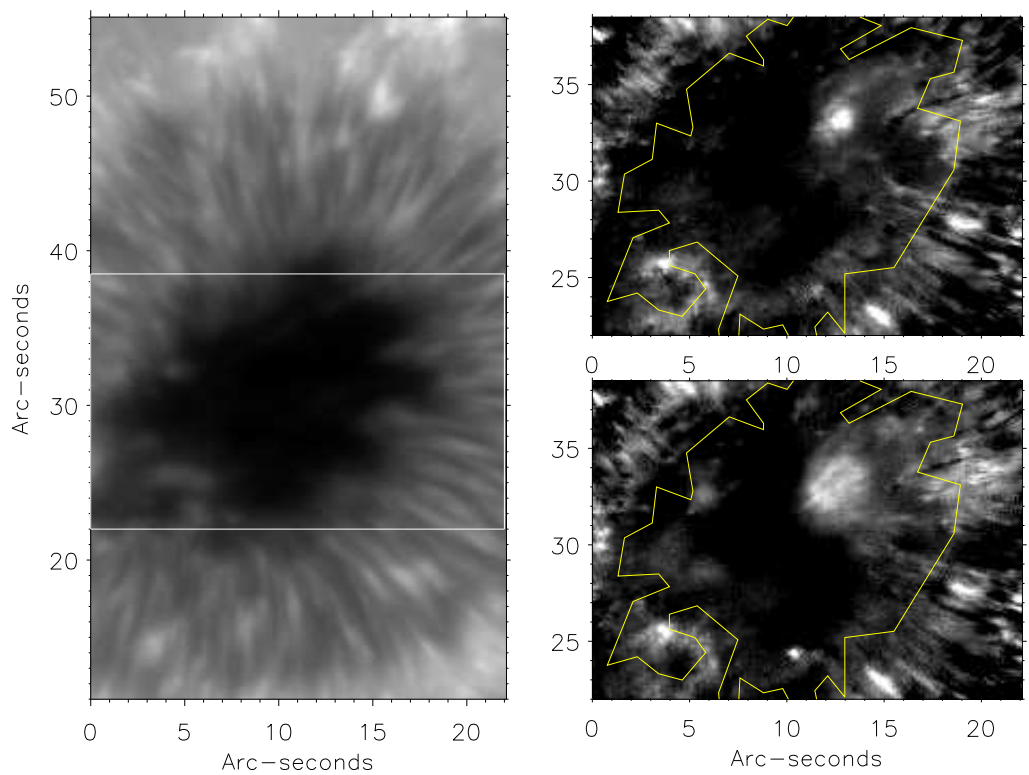}
\caption{Left: Field of view analyzed in the right panel. Right:
  Two successive snapshots showing the propagation of an umbral flash
  forming a ring-like structure around the origin.
\label{fig:ring}
}
\end{figure*}

There was one instance where we detected a bright flash that started
as a compact emission and then propagated outwards forming a ring (see
Fig~\ref{fig:ring}). This flash appears to be near the center of one
of the umbral cores and perhaps what we see here is the propagation
along field lines fanning out. The propagation speed in this case is
33.5$\pm$3.5~km~s$^{-1}$, very close to what we had before for the
propagation along the filaments in Fig~\ref{fig:filaments}. The
difference with the Doppler velocity measured in flashes suggests
again that this is not an actual motion of material.

\begin{figure}
\epsscale{1}
\plotone{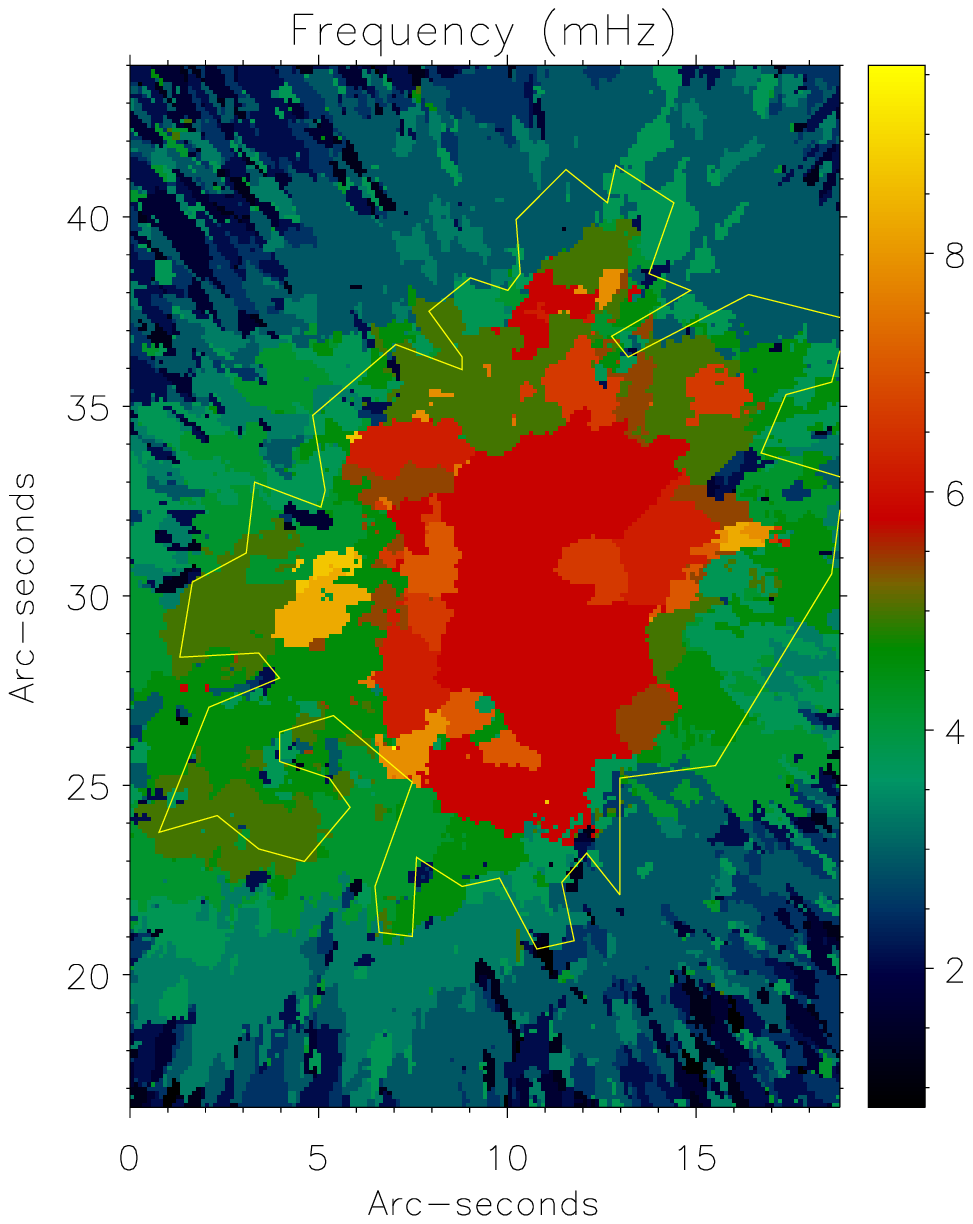}
\caption{Spatial distribution of the dominant oscillation
  frequency.
\label{fig:fourier}
}
\end{figure}

A Fourier analysis of the intensity oscillation shows that the umbra
does not oscillate as a whole (see Figure~\ref{fig:fourier}). Instead,
it is divided into smaller regions (resembling the umbral cores)
inside which the oscillation has a similar dominant frequency. This
property, however, can vary abruptly from one region to another. Such
subdivision of the umbra into smaller oscillating elements had already
been observed before (\citeNP{lites}; \citeNP{AV93}). Part of the
variation in the 
dominant frequency is likely a reflection of the gradient in field
inclination across the umbra (\citeNP{MIJ06}). 




\section{Conclusions}

We have presented new observations of the umbral chromosphere that
finally prove the existence of very fine structure in umbral flashes,
as had been anticipated by \citeN{sci} and \citeN{Cetal} based on
indirect spectroscopic evidence. Using Hinode's unique capabilities of
very high spatial resolution and time stability, it is now possible to
actually see such structure directly. Whether the dark component seen
here can be identified with the cool component at rest described in
those papers still remains to be confirmed.

To our surprise, we have found evidence for filamentary structure
inside the umbra with an important horizontal component (if not nearly
horizontal), which is difficult to understand in terms of the simple
cartoon picture of a theoretician's spot. Once more, while simple
idealized models are an excellent tool to understand global
properties, reality is also rich in intricate details that need to be
studied carefully as well. 

Clearly, the nature and properties of the filaments deserve further
study. Ideally one would like to have Stokes spectroscopy in order to
infer the actual orientation of the magnetic field. Unfortunately, no
instrument exists today that can do spectro-polarimetry of a \CaII \
line with the necessary spatial resolution ($\sim$0.2''). Hopefully, 2D
imaging filter-polarimeters such as IBIS (\citeNP{ibis}) or CRISP
(\citeNP{crisp}) may be able to 
accomplish this in the near future (but observing the infrared triplet,
not the H line). 

Perhaps even more mysterious is the existence of fine structure in the
chromospheric oscillation. One would think that it must be the
reflection of fine structure in the lower layers (possibly below the
photosphere) where the waves are excited. Numerical simulations
indicate that, in an umbra, waves must propagate upwards in the
vertical direction and are channeled by the field
lines~(\citeNP{bogdan}). The photospheric magnetic field appears even
at very high resolution to be very homogeneous and vertically
oriented. Therefore, the field cannot be responsible for the
inhomogeneities in the oscillation and the source must be in the
excitation mechanism itself.


\begin{thebibliography}{28}
\expandafter\ifx\csname natexlab\endcsname\relax\def\natexlab#1{#1}\fi

\bibitem[{{Aballe Villero} {et~al.}(1993){Aballe Villero}, {Marco}, {Vazquez},
  \& {Garcia de La Rosa}}]{AV93}
{Aballe Villero}, M.~A., {Marco}, E., {Vazquez}, M., \& {Garcia de La Rosa},
  J.~I. 1993, \aap, 267, 275

\bibitem[{{Beckers} \& {Tallant}(1969)}]{BT69}
{Beckers}, J.~M., \& {Tallant}, P.~E. 1969, \solphys, 7, 351

\bibitem[{{Bellot Rubio}(2007)}]{bellotreview}
{Bellot Rubio}, L.~R. 2007, in Highlights of Spanish Astrophysics IV, ed.
  F.~{Figueras}, J.~M. {Girart}, M.~{Hernanz}, \& C.~{Jordi}, 271--+

\bibitem[{{Bogdan} {et~al.}(2003){Bogdan}, {Carlsson}, {Hansteen}, {McMurry},
  {Rosenthal}, {Johnson}, {Petty-Powell}, {Zita}, {Stein}, {McIntosh}, \&
  {Nordlund}}]{bogdan}
{Bogdan}, T.~J., {Carlsson}, M., {Hansteen}, V.~H., {McMurry}, A., {Rosenthal},
  C.~S., {Johnson}, M., {Petty-Powell}, S., {Zita}, E.~J., {Stein}, R.~F.,
  {McIntosh}, S.~W., \& {Nordlund}, {\AA}. 2003, \apj, 599, 626

\bibitem[{{Carlsson} {et~al.}(2007){Carlsson}, {Hansteen}, {de Pontieu},
  {McIntosh}, {Tarbell}, {Shine}, {Tsuneta}, {Katsukawa}, {Ichimoto},
  {Suematsu}, {Shimizu}, \& {Nagata}}]{Carlsson}
{Carlsson}, M., {Hansteen}, V.~H., {de Pontieu}, B., {McIntosh}, S., {Tarbell},
  T.~D., {Shine}, D., {Tsuneta}, S., {Katsukawa}, Y., {Ichimoto}, K.,
  {Suematsu}, Y., {Shimizu}, T., \& {Nagata}, S. 2007, \pasj, 59, 663

\bibitem[{{Cauzzi} {et~al.}(2006){Cauzzi}, {Cavallini}, {Reardon}, {Berrilli},
  {Rimmele}, \& {IBIS Team}}]{ibis}
{Cauzzi}, G., {Cavallini}, F., {Reardon}, K., {Berrilli}, F., {Rimmele}, T., \&
  {IBIS Team}. 2006, in Bulletin of the American Astronomical Society, Vol.~38,
  Bulletin of the American Astronomical Society, 226--+

\bibitem[{{Centeno} {et~al.}(2005){Centeno}, {Socas-Navarro}, {Collados}, \&
  {Trujillo Bueno}}]{Cetal}
{Centeno}, R., {Socas-Navarro}, H., {Collados}, M., \& {Trujillo Bueno}, J.
  2005, \apj, 635, 670

\bibitem[{{Fontenla} {et~al.}(1990){Fontenla}, {Avrett}, \& {Loeser}}]{FAL}
{Fontenla}, J.~M., {Avrett}, E.~H., \& {Loeser}, R. 1990, \apj, 355, 700

\bibitem[{{Kneer} {et~al.}(1981){Kneer}, {Mattig}, \& {v.~Uexkuell}}]{UF8}
{Kneer}, F., {Mattig}, W., \& {v.~Uexkuell}, M. 1981, \aap, 102, 147

\bibitem[{{Kosugi} {et~al.}(2007){Kosugi}, {Matsuzaki}, {Sakao}, {Shimizu},
  {Sone}, {Tachikawa}, {Hashimoto}, {Minesugi}, {Ohnishi}, {Yamada}, {Tsuneta},
  {Hara}, {Ichimoto}, {Suematsu}, {Shimojo}, {Watanabe}, {Shimada}, {Davis},
  {Hill}, {Owens}, {Title}, {Culhane}, {Harra}, {Doschek}, \&
  {Golub}}]{Hinode1}
{Kosugi}, T., {Matsuzaki}, K., {Sakao}, T., {Shimizu}, T., {Sone}, Y.,
  {Tachikawa}, S., {Hashimoto}, T., {Minesugi}, K., {Ohnishi}, A., {Yamada},
  T., {Tsuneta}, S., {Hara}, H., {Ichimoto}, K., {Suematsu}, Y., {Shimojo}, M.,
  {Watanabe}, T., {Shimada}, S., {Davis}, J.~M., {Hill}, L.~D., {Owens}, J.~K.,
  {Title}, A.~M., {Culhane}, J.~L., {Harra}, L.~K., {Doschek}, G.~A., \&
  {Golub}, L. 2007, \solphys, 243, 3

\bibitem[{{Lites}(1986)}]{lites}
{Lites}, B.~W. 1986, \apj, 301, 992

\bibitem[{{L{\'o}pez Ariste} {et~al.}(2001){L{\'o}pez Ariste}, {Socas-Navarro},
  \& {Molodij}}]{UF5}
{L{\'o}pez Ariste}, A., {Socas-Navarro}, H., \& {Molodij}, G. 2001, \apj, 552,
  871

\bibitem[{{McIntosh} \& {Jefferies}(2006)}]{MIJ06}
{McIntosh}, S.~W., \& {Jefferies}, S.~M. 2006, \apjl, 647, L77

\bibitem[{{Nagashima} {et~al.}(2007){Nagashima}, {Sekii}, {Kosovichev},
  {Shibahashi}, {Tsuneta}, {Ichimoto}, {Katsukawa}, {Lites}, {Nagata},
  {Shimizu}, {Shine}, {Suematsu}, {Tarbell}, \& {Title}}]{UF1}
{Nagashima}, K., {Sekii}, T., {Kosovichev}, A.~G., {Shibahashi}, H., {Tsuneta},
  S., {Ichimoto}, K., {Katsukawa}, Y., {Lites}, B., {Nagata}, S., {Shimizu},
  T., {Shine}, R.~A., {Suematsu}, Y., {Tarbell}, T.~D., \& {Title}, A.~M. 2007,
  \pasj, 59, 631

\bibitem[{{Nye} \& {Thomas}(1974)}]{NT74}
{Nye}, A.~H., \& {Thomas}, J.~H. 1974, \solphys, 38, 399

\bibitem[{{Rouppe van der Voort} {et~al.}(2003){Rouppe van der Voort},
  {Rutten}, {S{\"u}tterlin}, {Sloover}, \& {Krijger}}]{VdV03}
{Rouppe van der Voort}, L.~H.~M., {Rutten}, R.~J., {S{\"u}tterlin}, P.,
  {Sloover}, P.~J., \& {Krijger}, J.~M. 2003, \aap, 403, 277

\bibitem[{{Scharmer} {et~al.}(2007){Scharmer}, {Langhans}, {Kiselman}, \&
  {L{\"o}fdahl}}]{scharmeretalreview}
{Scharmer}, G.~B., {Langhans}, K., {Kiselman}, D., \& {L{\"o}fdahl}, M.~G.
  2007, in Astronomical Society of the Pacific Conference Series, Vol. 369, New
  Solar Physics with Solar-B Mission, ed. K.~{Shibata}, S.~{Nagata}, \&
  T.~{Sakurai}, 71--+

\bibitem[{{Scharmer} {et~al.}(2008){Scharmer}, {Narayan}, {Hillberg}, {de la
  Cruz Rodriguez}, {Lofdahl}, {Kiselman}, {Sutterlin}, {van Noort}, \&
  {Lagg}}]{crisp}
{Scharmer}, G.~B., {Narayan}, G., {Hillberg}, T., {de la Cruz Rodriguez}, J.,
  {Lofdahl}, M.~G., {Kiselman}, D., {Sutterlin}, P., {van Noort}, M., \&
  {Lagg}, A. 2008, ArXiv e-prints

\bibitem[{{Sch{\"u}ssler} \& {V{\"o}gler}(2006)}]{magnetoconvection}
{Sch{\"u}ssler}, M., \& {V{\"o}gler}, A. 2006, \apjl, 641, L73

\bibitem[{{Shimizu} {et~al.}(2008){Shimizu}, {Nagata}, {Tsuneta}, {Tarbell},
  {Edwards}, {Shine}, {Hoffmann}, {Thomas}, {Sour}, {Rehse}, {Ito},
  {Kashiwagi}, {Tabata}, {Kodeki}, {Nagase}, {Matsuzaki}, {Kobayashi},
  {Ichimoto}, \& {Suematsu}}]{Hinode4}
{Shimizu}, T., {Nagata}, S., {Tsuneta}, S., {Tarbell}, T., {Edwards}, C.,
  {Shine}, R., {Hoffmann}, C., {Thomas}, E., {Sour}, S., {Rehse}, R., {Ito},
  O., {Kashiwagi}, Y., {Tabata}, M., {Kodeki}, K., {Nagase}, M., {Matsuzaki},
  K., {Kobayashi}, K., {Ichimoto}, K., \& {Suematsu}, Y. 2008, \solphys, 249,
  221

\bibitem[{{Socas-Navarro} {et~al.}(2000{\natexlab{a}}){Socas-Navarro},
  {Trujillo Bueno}, \& {Ruiz Cobo}}]{SNRCTB01}
{Socas-Navarro}, H., {Trujillo Bueno}, J., \& {Ruiz Cobo}, B.
  2000{\natexlab{a}}, \apj, 544, 1141

\bibitem[{{Socas-Navarro} {et~al.}(2000{\natexlab{b}}){Socas-Navarro},
  {Trujillo Bueno}, \& {Ruiz Cobo}}]{sci}
---. 2000{\natexlab{b}}, Science, 288, 1396

\bibitem[{{Socas-Navarro} {et~al.}(2001){Socas-Navarro}, {Trujillo Bueno}, \&
  {Ruiz Cobo}}]{SN01}
---. 2001, \apj, 550, 1102

\bibitem[{{Tsuneta} {et~al.}(2008){Tsuneta}, {Ichimoto}, {Katsukawa}, {Nagata},
  {Otsubo}, {Shimizu}, {Suematsu}, {Nakagiri}, {Noguchi}, {Tarbell}, {Title},
  {Shine}, {Rosenberg}, {Hoffmann}, {Jurcevich}, {Kushner}, {Levay}, {Lites},
  {Elmore}, {Matsushita}, {Kawaguchi}, {Saito}, {Mikami}, {Hill}, \&
  {Owens}}]{hinode}
{Tsuneta}, S., {Ichimoto}, K., {Katsukawa}, Y., {Nagata}, S., {Otsubo}, M.,
  {Shimizu}, T., {Suematsu}, Y., {Nakagiri}, M., {Noguchi}, M., {Tarbell}, T.,
  {Title}, A., {Shine}, R., {Rosenberg}, W., {Hoffmann}, C., {Jurcevich}, B.,
  {Kushner}, G., {Levay}, M., {Lites}, B., {Elmore}, D., {Matsushita}, T.,
  {Kawaguchi}, N., {Saito}, H., {Mikami}, I., {Hill}, L.~D., \& {Owens}, J.~K.
  2008, \solphys, 249, 167

\bibitem[{{Turova} {et~al.}(1983){Turova}, {Teplitskaia}, \& {Kuklin}}]{UF7}
{Turova}, I.~P., {Teplitskaia}, R.~B., \& {Kuklin}, G.~V. 1983, \solphys, 87, 7

\bibitem[{{Tziotziou} {et~al.}(2007){Tziotziou}, {Tsiropoula}, {Mein}, \&
  {Mein}}]{UF2}
{Tziotziou}, K., {Tsiropoula}, G., {Mein}, N., \& {Mein}, P. 2007, \aap, 463,
  1153

\bibitem[{{Tziotziou} {et~al.}(2002){Tziotziou}, {Tsiropoula}, \& {Mein}}]{UF4}
{Tziotziou}, K., {Tsiropoula}, G., \& {Mein}, P. 2002, \aap, 381, 279

\bibitem[{{Wittmann}(1969)}]{W69}
{Wittmann}, A. 1969, \solphys, 7, 366

\end{thebibliography}

\acknowledgments

The observations used in this paper were planned by J. Jurc{\'a}k and
Y. Katsukawa. Hinode is a Japanese mission developed and launched by
ISAS/JAXA, collaborating with NAOJ as a domestic partner, NASA and
STFC (UK) as international partners. Scientific operation of the
Hinode mission is conducted by the Hinode science team organized at
ISAS/JAXA. This team mainly consists of scientists from institutes in
the partner countries. Support for the post-launch operation is
provided by JAXA and NAOJ (Japan), STFC (U.K.), NASA, ESA, and NSC
(Norway).

\end{document}